\documentclass[twoside]{article}
\usepackage{fleqn,espcrc2,epsfig}

\usepackage{graphicx}
\usepackage[figuresright]{rotating}

\newcommand{\AmS}{{\protect\the\textfont2
  A\kern-.1667em\lower.5ex\hbox{M}\kern-.125emS}}

\title{Experimental Status of Semileptonic $B$ Decays \\
{\large Recent Results from LEP and Comparisons with $\Upsilon (4S)$ 
Experiments}}

\author{M. Battaglia\address{Dept. of Physics,
        University of Helsinki, \\ 
        FIN-00014 Helsinki, Finland}%
        \thanks{present address: CERN/EP Division, CH-1211 Geneva}}
       
\begin{document}

\begin{abstract}

Recent analyses of the LEP and $\Upsilon (4S)$ data have better outlined the
picture of semileptonic $B$ decays. Results on inclusive and exclusive decay
branching fractions and on the extraction of the $|V_{ub}|$ and $|V_{cb}|$ 
elements of the CKM mixing matrix are discussed, together with some of the 
still open questions and the sources of model systematics.
\vspace{1pc}
\end{abstract}

\maketitle

\section{INTRODUCTION}

Semileptonic {(s.l.)} $B$ decays represent a favorable laboratory to study
the dynamics of heavy quark decays, to determine the size of the $|V_{ub}|$ 
and $|V_{cb}|$ CKM mixing matrix elements, crucial in the unitarity triangle 
test of the Standard Model, and to acquire informations on the structure of the
$B$ meson itself.

The {\sc Aleph}, {\sc Delphi}, L3 and {\sc Opal} experiments have analyzed 
together about 1.5~M s.l.\ $B$ decays recorded at {\sc Lep}, at energies 
around 
the $Z^0$ pole from 1990 to 1995. This statistics is significantly lower
than about 4~M s.l.\ decays recorded by {\sc Cleo} and 7.5~M already
logged by the {\sc BaBar} and {\sc Belle} experiments. However, the 
significant boost of the beauty hadrons, the confinement of their decay 
products in well separated jets and the production of almost all beauty hadron
species make it possible to perform topological decay reconstruction with good
efficiency on the {\sc Lep} data sets, thus exploiting analysis techniques 
complementary to those used at lower energy $e^+e^-$ colliders. Since operator
product expansion (OPE) predictions apply to sufficiently 
inclusive observables, it is advantageous to reconstruct s.l.\ decays 
inclusively, with only mild cuts on the lepton and hadron energies.
Further, the {\sc Lep} kinematics allow the two extreme kinematical regions at
small and large momentum transfer to be accessed, since the $B$ decay products
gain enough energy from the $B$ boost.

Section~2 summarizes recent results on inclusive and exclusive s.l.\ 
branching fractions obtained by the {\sc Lep} experiments. These results have
been averaged and compared with those from experiments running at the 
$\Upsilon (4S)$. Section~3 
discusses the extraction of the $|V_{ub}|$ and $|V_{cb}|$ elements from 
inclusive and exclusive s.l.\ yields. Section~4 reviews some open 
questions in s.l.\ decays and model systematics presently limiting the 
precision in the interpretation of the data.

\section{RECENT RESULTS}

\subsection{Inclusive s.l.\ Branching Fraction}

The determination of the inclusive s.l.\ branching fraction 
BR($b \rightarrow X \ell \bar \nu$) is important for the measurement of
$|V_{cb}|$. Together with the measurement of the charm multiplicity and of 
exclusive decays discussed later in this Section, they provide with a test of 
the s.l.\ decay width.
The main experimental issue here is the separation of the prompt 
$b \rightarrow \ell$ signal from the cascade $b \rightarrow c (\bar c) 
\rightarrow \bar \ell (\ell)$ and the $c \rightarrow \bar \ell$ backgrounds. 
This has been usually achieved by performing a fit to the lepton $p$ and $p_t$ 
distributions to profit from the harder spectra of the prompt signal.
However, this method is limited by the model dependence in the description 
of the prompt and cascade spectra, and alternative techniques have recently 
been exploited to improve the signal separation in a more model-independent 
way. 
These analyses use the charge correlation between the $b$ and the lepton, the 
decay topology, and double tagged events where both beauty hadrons decay 
semileptonically. The dominant source of uncertainty remains, however, that 
due to the modeling of the $b \rightarrow \ell$ and $c \rightarrow \bar \ell$ 
spectra.
A global fit to the {\sc Lep} and {\sc Sld} electro-weak data, including also 
$R_b$, $R_c$ and asymmetry determinations by analyses using leptons and other 
independent methods, constrains the s.l.\ decay models, thus reducing the 
related systematics by about 20\%~\cite{lepewwg}. However, this depends on the
relative size of the systematic uncertainties assigned to different asymmetry
measurements, and a more conservative approach is to derive the {\sc Lep} 
average from the direct determinations. This gives 
BR($b \rightarrow X \ell \bar \nu$) = 0.1056 $\pm$ 0.0011~(stat) $\pm$ 
0.0018~(syst). After having rescaled this value by $\frac{1/2 (\tau(B_d) + 
\tau(B_u))}{\tau(b)} = 1.021 \pm 0.013$, to account for the different beauty 
hadron species produced, this result can be compared with the most recent 
determination at the $\Upsilon (4S)$, obtained by {\sc Cleo} using a lepton 
tagged method to separate the prompt lepton yield in $B$ decays from the 
backgrounds giving BR($b \rightarrow X \ell \bar \nu$) = 0.1049 $\pm$ 
0.0017~(stat)  $\pm$ 0.0043~(syst)~\cite{cleo_sl}.

It is useful to analyze these results in relation with
the average number of charm particles in $B$ decays, $n_c$, since the
semileptonic, double charmed and charmless yields are correlated once the 
total decay width is fixed. There are three main sources of data on $n_c$. 
The results and their average are summarized in Table~\ref{tab:nc}.
The first is based on the
determination of the $B$ decay branching ratios into individual charm hadrons
BR($B \rightarrow D^{+}, D^0, D^+_s, \Lambda^+_c + X$) and their
anti-particles. 

\vspace{-0.5cm}

\begin{table}[htb]  
\caption[]{\sl Determinations of $n_c$ at {\sc Lep} and by {\sc Cleo}.}
             
\begin{center}
\begin{tabular}{c c c}
\hline 
Method & {\sc Lep} & {\sc Cleo} \\
\hline
Charm counting & 1.178 $\pm$ 0.074 & 1.132 $\pm$ 0.062\\
Wrong sign $D$ & 1.293 $\pm$ 0.075 & 1.174 $\pm$ 0.053\\
Topology & 1.151 $\pm$ 0.047  & - \\ \hline
Average & 1.171 $\pm$ 0.040   & 1.159 $\pm$ 0.049 \\
\hline
\end{tabular}
\end{center}
\label{tab:nc}

\vspace{-0.25cm}

\end{table}

These rates have been determined at {\sc Lep} and by {\sc Cleo} and need to be
further corrected to include the strange charm baryon and charmonium
production in $B$ decays. A second approach is to express $n_c$ as
$n_c = 1 + 2 \times {\mathrm{BR}}(b \rightarrow c \bar c s) - 
{\mathrm{BR}}(b \rightarrow {\mathrm {charmless}})$
and to determine BR($b \rightarrow c \bar c s$) more inclusively from 
BR($b \rightarrow J/\Psi X$) and BR($b \rightarrow D \bar D X$).
There have been several recent results on the decay branching fractions into
open double charm $D \bar D X$ or wrong sign $D (\bar D X)$ at {\sc Lep}, 
by {\sc Cleo} and also by {\sc BaBar}. Finally a third method, exploited 
by {\sc Delphi} and {\sc Sld}, consists in extracting the double charm and 
charmless  yield from the $B$ decay topology. 
\begin{figure}[htb]

\vspace{-0.5cm}

\epsfig{file=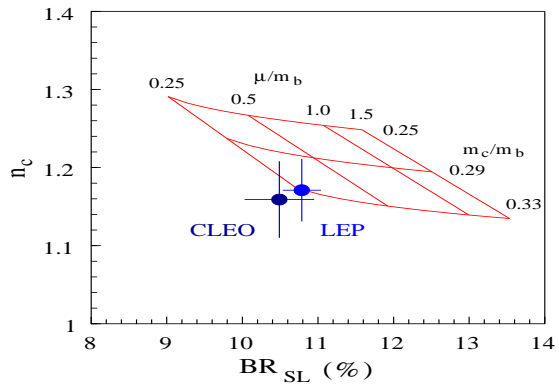,width=7.25cm,height=5.0cm}

\vspace{-1.0cm}

\caption{\sl Present results for the average number of charm hadrons in 
$B$ decays $n_c$ and inclusive s.l.\ branching fraction BR$_{sl}$ from 
{\sc Lep} and {\sc Cleo} compared with the predictions of 
Ref.~\cite{neubert1}.}
\label{fig:brslnc}

\vspace{-0.5cm}

\end{figure}
The scaled {\sc Lep} averages and the {\sc Cleo} 
values can be independently compared with theoretical predictions, obtained
using heavy quark expansion to order $1/m_b^2$. These are shown in 
Figure~\ref{fig:brslnc} for a range of values of the renormalization scale 
$\mu$ and of the ratio of the quark pole masses 
$m_c/m_b$~\cite{neubert1}. The experimental results agree with these 
predictions at $\mu \simeq 0.25$ and $m_c/m_b \simeq 0.33$.

\subsection{Charmless s.l.\ Branching Fraction}

An accurate determination of the charmless s.l.\ $B$ branching fraction is
important for the measurement of the $|V_{ub}|$ element and has represented 
a significant challenge to both theorists and experiments since {\sc Cleo} 
first observed charmless s.l.\ decays as an excess of leptons at 
energies above the kinematical limit for decays with an accompanying charm 
hadron~\cite{cleo1}. 
However, the limited fraction ($\simeq$ 10\%) of charmless decays
populating this end-point region results in a significant model uncertainty 
for the extraction of $|V_{ub}|$. An analysis technique based on the invariant
mass $M_X$ of the hadronic system recoiling against the lepton pair, peaked 
for $b \rightarrow X_u \ell \bar \nu$ at a significantly lower value than 
for $b \rightarrow X_c \ell \bar \nu$, was proposed several years 
ago~\cite{barger} and it has been the subject of new theoretical 
calculations~\cite{falk,bdu,defazio1}. If $b \rightarrow u$ 
transitions can be discriminated from the dominant $b \rightarrow c$ 
background up to $M_X \simeq M(D)$, this method is sensitive to 
$\simeq$~80\% of the charmless s.l.\ $B$ decay rate. Further, if no 
preferential weight is given to low mass states in the event selection, the 
non-perturbative effects are expected to be small and the OPE description of 
the transition has been shown to be accurate away from the resonance region. 
The requirement to isolate the $b \rightarrow u$ contribution to the s.l.\ 
yield from the $\simeq$~60 times larger $b \rightarrow c$ one, while ensuring 
an uniform sampling of the decay phase space to avoid biases towards the few 
exclusive low-mass, low-multiplicity states, make this analysis a major 
experimental challenge. 
\begin{figure}[ht!]

\vspace{-0.5cm}

\begin{center}
\epsfig{file=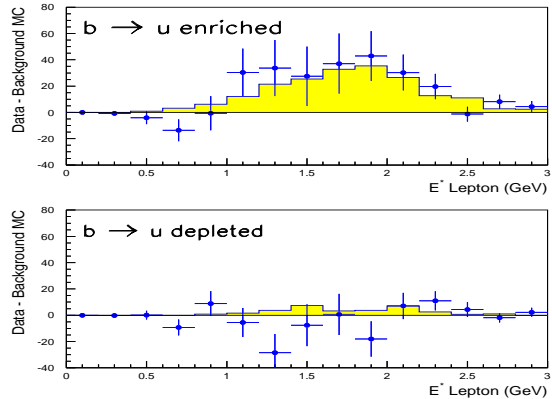,width=8.0cm,height=6.0cm}
\end{center}

\vspace{-1.25cm}

\caption[]{\sl Background subtracted distributions for the lepton energy in 
the $B$ rest frame $E^*_{\ell}$ obtained in the {\sc Delphi} analysis: the 
$b \rightarrow u$ enriched decays with $M_X <$ 1.6~GeV/c$^2$ (upper plot)
and $b \rightarrow u$ depleted decays with $M_X <$ 1.6~GeV/c$^2$ (lower plot).
The shaded histograms show the expected $E^*_{\ell}$ distribution for signal 
$b \rightarrow u$ s.l.\ decays.}
\label{fig:delphivub}

\vspace{-0.5cm}

\end{figure}
{\sc Aleph}~\cite{aleph_vub}, {\sc Delphi}~\cite{delphi_vub} and 
L3~\cite{l3_vub} have developed new analysis techniques based 
on the observation that $b \rightarrow X_u \ell \bar \nu$ decays can be 
inclusively selected from $b \rightarrow X_c \ell \bar \nu$ by
the difference in the invariant mass and kaon content of the 
secondary hadronic system and in the decay multiplicity and vertex topology.
These features have been implemented differently in the analyses by the three 
experiments: {\sc Aleph} used a neural net discriminant based on kinematical 
variables, {\sc Delphi} preferred a classification of s.l.\ decays on the basis
of their reconstructed hadronic mass $M_X$, decay topology and presence of 
secondary kaons and L3 adopted a sequential cut analysis based on the 
kinematics of the two leading hadrons in the same hemisphere as the tagged 
lepton. Starting from a natural signal-to-background ratio S/B of about 
0.02, {\sc Aleph} obtained S/B = 0.07 with an efficiency $\epsilon =$ 11\%, 
{\sc Delphi} had S/B = 0.10 with $\epsilon =$ 6.5\% and L3 had S/B~=~0.16 with 
$\epsilon =$ 1.5\%.
 
\begin{table*}[htb]
\caption[]{\sl Summary of the {\sc Lep} BR($b \rightarrow X_u \ell \bar \nu$) 
results with the sources of the statistical, experimental, uncorrelated and 
correlated systematic uncertainties.}             

\begin{center}
\begin{tabular}{@{}lccccc}
\hline
Experiment & BR($b \rightarrow X_u \ell \bar \nu$) & (stat.) & (exp.) & 
(uncorrelated) & (correlated) \\
\hline \hline
& & & & &\\
{\sc Aleph}~\cite{aleph_vub} & (1.73 & $\pm$ 0.48 & $\pm$ 0.29 &
$\pm$ 0.29 ($^{\pm 0.29~b\rightarrow c}_{\pm 0.08~b\rightarrow u}$) & 
$\pm$ 0.47 ($^{\pm 0.43~b\rightarrow c}_{\pm 0.19~b\rightarrow u}$))
$\times 10^{-3}$ \\
 & & & & &\\ 
{\sc Delphi}~\cite{delphi_vub} & (1.69 & $\pm$ 0.37 & $\pm$ 0.39 &
$\pm$ 0.18 ($^{\pm 0.13~b\rightarrow c}_{\pm 0.13~b\rightarrow u}$) 
& $\pm$ 0.42 ($^{\pm 0.34~b\rightarrow c}_{\pm 0.20~b\rightarrow u}$))
$\times 10^{-3}$ \\
& & & & &\\
{\sc L3}~\cite{l3_vub} & (3.30 & $\pm$ 1.00 & $\pm$ 0.80 & 
$\pm$ 0.68 ($^{\pm 0.68~b\rightarrow c}_{~~~-~~b\rightarrow u}$)
& $\pm$ 1.40 ($^{\pm 1.29~b\rightarrow c}_{\pm 0.54~b\rightarrow u}$))
$\times 10^{-3}$ \\
& & & & &\\
\hline
\end{tabular}
\end{center}
\label{tab:summary}
\end{table*}
All three experiments observed a significant data excess 
over the estimated backgrounds corresponding to 
303 $\pm$ 88 events for {\sc Aleph}, 214 $\pm$ 56 for {\sc Delphi} and 
81 $\pm$ 25 for L3. The characteristics of these events correspond to those 
expected for $b \rightarrow X_u\ell \bar \nu$ 
decays~(see Figure~\ref{fig:delphivub}). The inclusive charmless s.l.\ 
branching ratios summarized in Table~~\ref{tab:summary} were obtained. 
These results are compatible within their uncorrelated uncertainties. 
The differences in the analysis techniques resulted in systematic 
uncertainties that are only partially correlated. The {\sc Lep} average value 
was therefore computed, resulting in BR($b \rightarrow X_u \ell \bar 
\nu$) = (1.74 $\pm$ 0.37 (stat.+exp.) $\pm$ 0.38 ($b \rightarrow c$) 
$\pm$ 0.21 ($b \rightarrow u$)) $\times$ 10$^{-3}$ =
(1.74 $\pm$ 0.57) $\times$ 10$^{-3}$~\cite{beach2000}. The {\sc Lep} 
experiments have shown the feasibility of these inclusive analyses, due to
the favorable kinematics and the decay reconstruction capabilities of their
detectors. The exact size of the systematic uncertainties attached to the 
inclusive analyses is still being debated within the community. 
While more precise data on $B$ and $D$ decays will decrease the dominant
$b \rightarrow c$ systematics of this measurement, future perspectives for 
inclusive charmless s.l.\ rate determinations are not yet clearly outlined.
The hadronic mass analysis puts even further problems to symmetric and 
asymmetric $B$ factories at the $\Upsilon(4S)$ peak, due to the confusion 
between the decay products of the $B$ and $\bar B$ and of the reduced $B$ 
decay length, compared with {\sc Lep}. 
A feasibility study, performed by {\sc BaBar}, 
requires full reconstruction of one $B$ meson through an exclusive decay mode 
to solve the first problem, and predicts a reconstructed signal sample with 
$M_X <$ 1.7~GeV/$c^2$ of about 100 to 300 events for the full data statistics 
of 100~fb$^{-1}$~\cite{babook}. A new technique based on the reconstruction
of the di-lepton $\ell \bar \nu$ invariant mass of the decay has been proposed
recently~\cite{ligeti1}. While this method is sensitive to $\simeq 20\%$ of 
the inclusive charmless rate, above the $b \rightarrow c$ threshold, it should 
provide a good control of theoretical uncertainties. At low-energy $B$ 
factories, the di-lepton mass measurement will profit from the $\nu$
reconstruction techniques that rely on the $B$ production at threshold, 
already successfully exploited by {\sc Cleo}.
 
\subsection{Exclusive s.l.\ $B$ Meson Decays}

Exclusive s.l.\ $B$ decays have been studied at the $\Upsilon(4S)$ and 
at {\sc Lep}, to establish the relative contribution of the individual
channels to the s.l.\ decay width and to extract the relevant CKM elements.
Charmless $B \rightarrow \pi \ell \nu$ and $\rho \ell \nu$ decays have been
observed by {\sc Cleo} and their rates measured~\cite{cleo2}. 
Candidates for these exclusive decays have also been isolated by 
{\sc Aleph}~\cite{aleph_vub} and {\sc Delphi}~\cite{delphi_vub}, from the 
inclusive analyses, but their statistics are insufficient for a rate 
determination with interesting accuracy. 

\begin{figure}[htb]

\vspace{-0.5cm}

\begin{center}
\epsfig{file=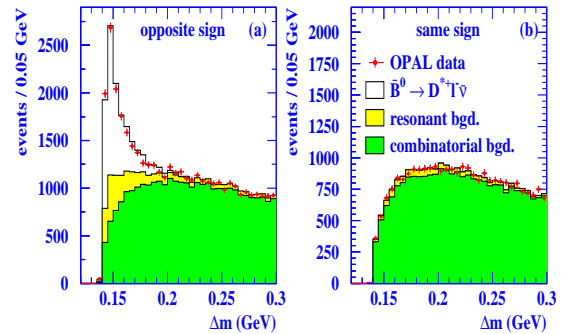,width=8.0cm,height=5.5cm}
\end{center}

\vspace{-1.25cm}

\caption[]{\sl The reconstructed $\Delta M = M(D \pi) - M(D)$ distribution for 
selected opposite (left) and same (right) sign lepton - pion combinations 
for the {\sc Opal} analysis. The $D^{*+}$ signal is visible in the opposite 
sign sample. In this inclusive analysis $M(D)$ is approximated by the invariant
mass of the partially reconstructed $D$ decay.}

\label{fig:opalvcb}

\vspace{-0.5cm}

\end{figure}

Competitive accuracies have been obtained at {\sc Lep} for charmed exclusive
decay modes. The decay $\bar{B^0_d} \rightarrow D^{*+} \ell^-
\bar \nu$ has received specific attention for the extraction of $|V_{cb}|$ from
a study of its rate $\frac {d \Gamma}{dw} = K(w) F^2(w) |V_{cb}|^2$ as a
function of the the $D^{*}$ and $B_d$ four-velocity product $w$. In this 
expression $K(w)$ is a phase space factor and $F(w)$ the hadronic form factor.
The reconstruction of the decay can be performed fully exclusively, through 
the decay $D^{*+} \rightarrow D^0 \pi^+$ followed by $D^0 \rightarrow K^- 
\pi^+$ and, at {\sc Lep}, also by partial inclusive reconstruction of the $D$ 
meson (see Figure~\ref{fig:opalvcb}) resulting in a significant increase in 
the selection efficiency. 
\begin{figure}[ht!]

\vspace{-0.5cm}

\begin{center}
\begin{tabular} {c c}
\epsfig{file=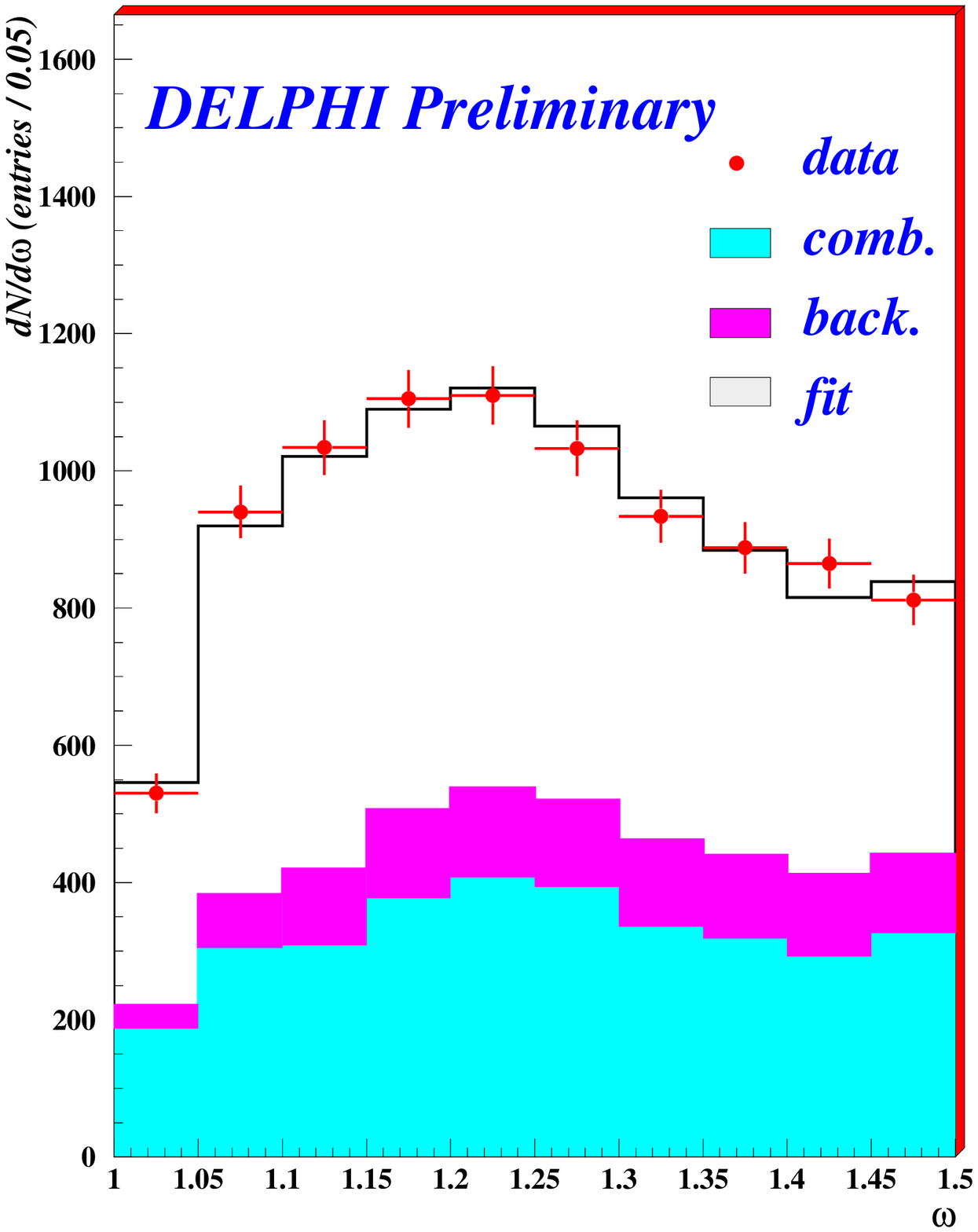,width=3.5cm,height=5.75cm} &
\epsfig{file=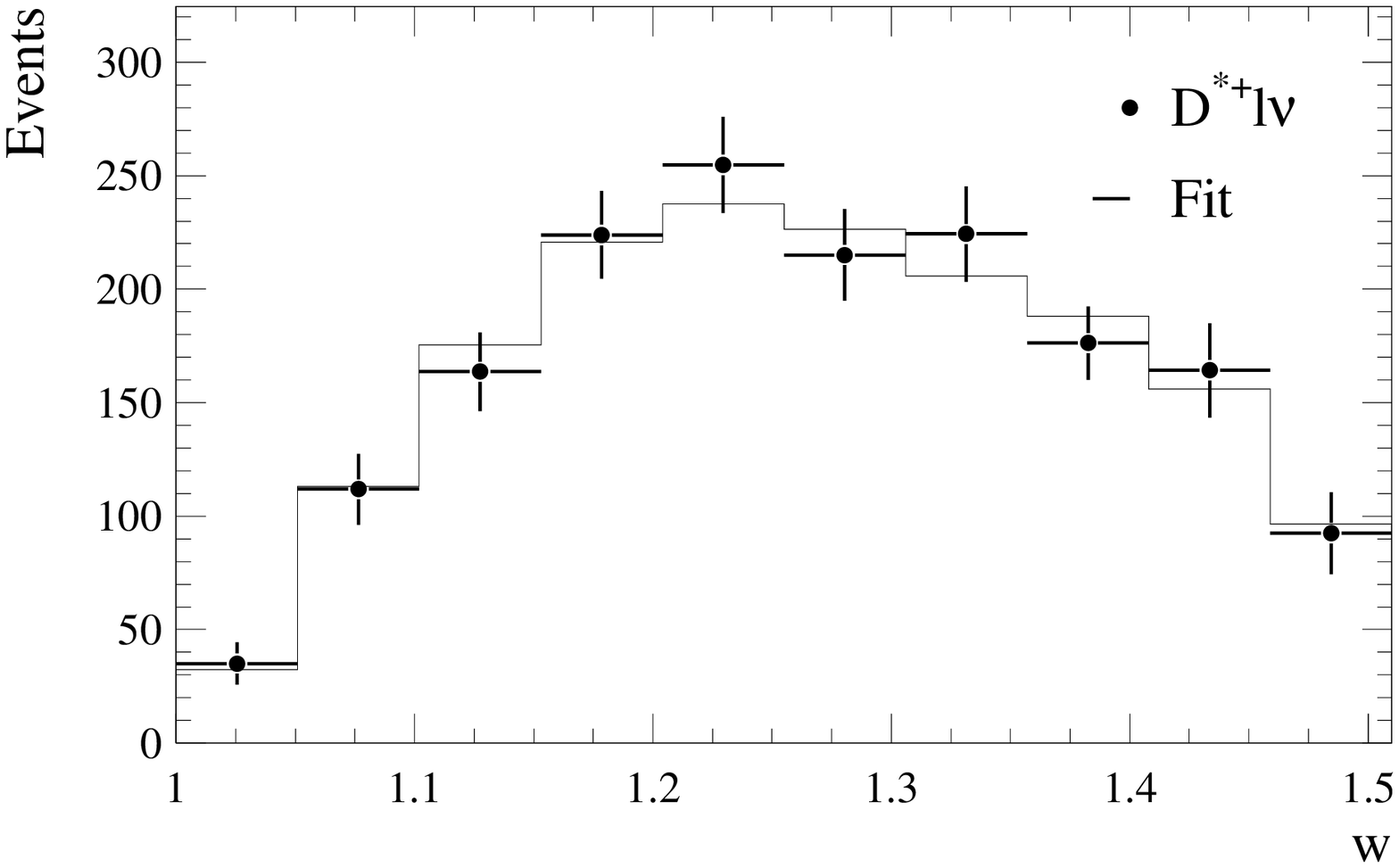,bbllx=15,bblly=10,bburx=567,bbury=360,width=3.5cm,
height=5.75cm,clip=} \\
\end{tabular}
\end{center}

\vspace{-1.25cm}

\caption[]{\sl The differential rate $dN/dw$ for $\bar{B^0_d} \rightarrow 
D^{*+} \ell^- \bar \nu$ events for the {\sc Delphi} inclusive analysis (left) 
and the {\sc Cleo} exclusive analysis (right).}

\label{fig:dndw}

\vspace{-0.5cm}

\end{figure}
At {\sc Lep}, decays can be efficiently reconstructed closer to zero 
$D^{*+}$ recoil energy than at the $\Upsilon(4S)$. But using exclusive 
reconstruction and at the $\Upsilon(4S)$, where the $B$ meson is almost at 
rest, better resolution on $w$ is achieved:
{\sc Cleo} obtains $\sigma(w)$ = 0.03 compared with $\sigma(w)$ = 0.07 to 
0.12 from the inclusive {\sc Lep} analyses (see Figure~\ref{fig:dndw}).  
\begin{figure}[h!]

\vspace{-0.5cm}

\begin{center}
\epsfig{file=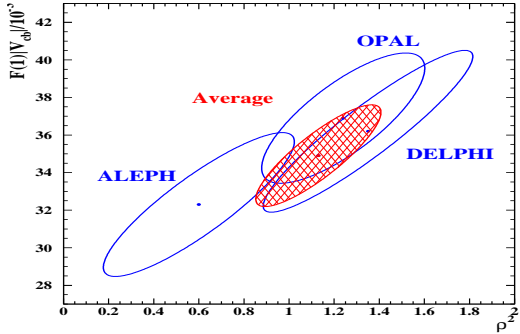,width=7.5cm,height=5.0cm}
\end{center}

\vspace{-1.25cm}

\caption[]{\sl The $F(1) |V_{cb}|$ and $\rho^2$ determinations at {\sc Lep}, 
corrected by the {\sc Lep} $|V_{cb}|$ Working Group for a consistent set 
of input parameters, and the resulting {\sc Lep} average. The ellipses indicate
the 65\% C.L. of each result.}
\label{fig:vcbrho}

\vspace{-0.5cm}

\end{figure}
The main background that these analysis have to reduce and understand
is due to s.l.\ transitions into charmed excited states $D^{**}$ producing a 
$D^{*}$ in their decay as discussed below.
The extrapolation of the form factor $F(w)$ is based on a dispersion-relation 
parameterization.
By combining the measurements by {\sc Aleph}, {\sc Delphi} and {\sc Opal}, the 
{\sc Lep} averages $F(1) |V_{cb}| = (34.5 \pm 0.7 ({\mathrm{(stat)}})
\pm 1.5 ({\mathrm{(syst)}})\times 10^{-3}$ and $\rho^2 = 1.13  \pm 0.08 
({\mathrm{(stat)}}) \pm 0.16 ({\mathrm{(syst)}})$ were obtained with a a fit
confidence level of 12\% (see Figure~\ref{fig:vcbrho}). 
{\sc Cleo} recently reported $F(1) |V_{cb}| = (42.4 \pm 1.8 ({\mathrm{(stat)}})
\pm 1.9 ({\mathrm{(syst)}}) \times 10^{-3}$ and 
$\rho^2 = 1.67  \pm 0.11 ({\mathrm{(stat)}}) \pm 0.22 ({\mathrm{(syst)}})$ that
gives 2.4~$\sigma$ higher extrapolated rate at $w=1$ with a larger 
slope~\cite{cleovcb}.
Since the intercept at $w=1$ and the slope are highly correlated, this prompts
further investigation of this decay through new analyses of the 
experimental data and of the underlying uncertainties.

\begin{figure}[hb!]

\vspace{-0.75cm}

\begin{center}
\epsfig{file=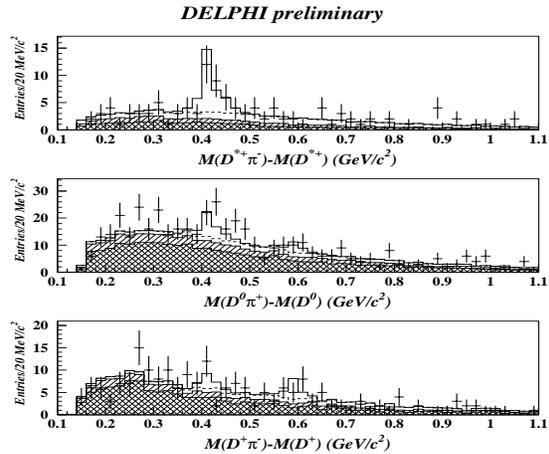,width=8.0cm,height=6.5cm}
\end{center}

\vspace{-1.25cm}

\caption[]{\sl The invariant mass difference in the decays 
$\bar B \rightarrow D^{*+} \pi^- \ell^- X$, $D^0 \pi^+ \ell^- X$ and 
$D^+ \pi^- \ell^- X$. The data are represented by the dots with error bars,
the solid open histogram is the result of a fit to the data including narrow
$D_1(2420)$ and $D^*_2(2460)$ states, broad $D^* \pi$ and $D \pi$ states
(dashed histogram), fake $D$ (cross-hatched histogram) and fragmentation
particle (hatched histogram) backgrounds.}
\label{fig:delphidpi}

\vspace{-0.5cm}

\end{figure}
The study of s.l.\ $B$ decays into orbitally and radially excited charmed
mesons has established the production of the narrow orbital excitation 
$D_1(2420)$, while the $D^*_2(2460)$ and a broad state decaying into 
$D^{*+} \pi^-$ have been observed by {\sc Cleo} in hadronic $B$ 
decays~\cite{cleo_broad}.
 
\begin{table*}[htb]
\caption[]{\sl Summary of the results on s.l.\ decays into excited charmed 
states.}
             
\begin{center}
\begin{tabular}{@{}lccc}
\hline
Experiment & BR($\bar B \rightarrow D_1(2420) \ell \bar \nu$) &
             BR($\bar B \rightarrow D^*_2(2460) \ell \bar \nu$) &
             BR($\bar B \rightarrow D^*_{(0,1)}({\mathrm{broad}}) \ell \bar 
\nu$) \\
\hline \hline
& & & \\
{\sc Aleph}~\cite{aleph_dpi} & (7.0 $\pm$ 1.1 $\pm$ 1.2) $\times 10^{-3}$ &
(2.4 $\pm$ 1.0 $\pm$ 0.5) $\times 10^{-3}$ &  \\
& & & \\ 
{\sc Delphi}~\cite{delphi_dpi} & (6.7 $\pm$ 1.8 $\pm$ 1.0) $\times 10^{-3}$ &
(4.4 $\pm$ 2.1 $\pm$ 1.2) $\times 10^{-3}$ &  
(22.9 $\pm$ 5.9 $\pm$ 3.6) $\times 10^{-3}$ \\
& & & \\
{\sc Cleo}~\cite{cleo_dpi} & (5.6 $\pm$ 1.3 $\pm$ 0.9) $\times 10^{-3}$ &
(3.0 $\pm$ 3.3 $\pm$ 0.8) $\times 10^{-3}$ &  \\
& & & \\
\hline
\end{tabular}
\end{center}
\label{tab:dpisum}
\end{table*}
A recent analysis by {\sc Delphi}~\cite{delphi_dpi} has separately measured
the narrow $D_1(2420)$ and broad, or non-resonant, $D^{(*)} \pi$ production 
(see Figure~\ref{fig:delphidpi}). 
Results are summarized in Table~\ref{tab:dpisum}.
The $D^*_2(2460)$ production rate is less than that of 
$D_1(2420)$:  $R^* = \frac {BR(\bar  B \rightarrow D_2 \ell^- \bar \nu)}
{BR(\bar  B \rightarrow D_1 \ell^- \bar \nu)} < 0.6$ in disagreement with 
the HQET prediction in the infinitely heavy quark mass limit $R^* \simeq 
1.6$~\cite{morenas}, thus implying large $1/m_Q$ corrections~\cite{ligeti2}.
Including these corrections restores agreement with the data for $R^*$, 
however 
the overall yield of the broad $D^*_0 + D^*_1$ states is not expected to 
exceed that for $D_1(2420)$. This apparent discrepancy with the preliminary 
{\sc Delphi} result can now be explained as an experimental effect, if the
feed-through of non-resonant $D^* \pi$ production~\cite{isgur} is non
negligible.

A first estimate of the slope of the $\Lambda_b$ form factor has also been 
obtained at {\sc Lep}. {\sc Delphi} performed a fit to the reconstructed $w$ 
distribution and event rate for a sample of candidate $\Lambda_b \rightarrow 
\Lambda_c \ell \bar \nu$ decays obtaining $\rho^2$ = 
1.55 $\pm$ 0.60~(stat) $\pm$ 0.55~(syst)~\cite{delphilb}, which is within the 
range of the current theoretical predictions.
 
\section{EXTRACTION OF $|V_{ub}|$ AND $|V_{cb}|$}

\subsection{$|V_{ub}|$}

The value of the $|V_{ub}|$ element can be extracted from the inclusive
charmless s.l.\ branching fraction BR($b \rightarrow X_u \ell \bar \nu$)
by using the  following relationship derived in the context of Heavy Quark 
Expansion~\cite{uraltsev,hoang}:
\begin{equation}
|V_{ub}| = .00445~(\frac{ \mathrm{BR}(b \rightarrow X_u \ell \bar \nu)}
{0.002} \frac{1.55 \mathrm{ps}}{\tau_b})^\frac{1}{2} \times
\end{equation}

\vspace{-0.5cm}

\begin{displaymath}
(1 \pm .020 \mathrm{(QCD)} \pm .035 \mathrm{(m_b)})
\end{displaymath}
\noindent where the value $m_b$ = (4.58 $\pm$ 0.06)~GeV/c$^2~$ has been 
assumed~\cite{lephfs}. The theoretical uncertainties are small, due to the
absence of $1/m_b$ term in the expansion and of $1/(m_b - m_c)$ 
dependence, and it is dominated by that on the $b$ mass. 
Inserting the {\sc Lep} average BR($b \rightarrow X_u \ell \bar \nu$), 
$|V_{ub}|$ was determined to be:
$|V_{ub}|~=~(4.13 ^{+0.42}_{-0.47} \mathrm{(stat.+det.)} 
      ^{+0.43}_{-0.48} \mathrm{(b \rightarrow c~syst.)}$
      $^{+0.24}_{-0.25} \mathrm{(b \rightarrow u~sys.)}$
      $\pm 0.02 (\tau_b) 
      \pm 0.20 \mathrm{(HQE)}) \times 10^{-3}$.
The $b \rightarrow u$ and HQE model systematics is slightly below 10\% and 
the large uncertainties from the modeling of $b \rightarrow c$ decays can be 
reduced in future by more precise data.
This inclusive determination of $|V_{ub}|$ can be compared to that extracted
from the determination of the exclusive rate for the decay
$B \rightarrow \rho \ell \bar \nu$ by {\sc Cleo} giving
$|V_{ub}|$ = (3.25 $\pm$ 0.14 (stat.) $^{+0.21}_{-0.29}$ (syst.)
$\pm$ 0.55 (model) ) $\times 10^{-3}$~\cite{cleo2}. 
The two measurements are consistent within their uncertainties, which are 
mostly uncorrelated. It is expected that the large model dependence of the
exclusive method will be reduced by computing the hadronic form factor 
from lattice QCD.

\subsection{$|V_{cb}|$}

There are two methods, to extract the $|V_{cb}|$ element from s.l.\ decays: 
i) from the inclusive s.l.\ $B$ decay width, once the charmless contribution 
has been subtracted, and ii) by the extrapolation of the rate for exclusive 
decays, as $B \rightarrow D^{*} \ell \bar \nu$, at zero recoil. The two 
results can be 
used as a check of the underlying theory and to improve the $|V_{cb}|$ 
accuracy by their averaging, the systematic uncertainties being partially 
uncorrelated.

For the inclusive method the relationship:
\begin{equation}
|V_{cb}| = .0411 \times (\frac{ \mathrm{BR}(b \rightarrow X_c \ell \nu)}
{.105} \frac{1.55 \mathrm{ps}}{\tau_b})^\frac{1}{2} \times
\end{equation}

\vspace{-0.5cm}

\begin{displaymath}
\big( 1 - 0.024 \times \big( \frac{\mu_{\pi}^2 - 0.5}{0.1} \big) \big) \\
\times (1 \pm 0.030 \mathrm{(pert.)}
\end{displaymath}

\vspace{-0.5cm}

\begin{displaymath}
\pm 0.020 \mathrm{(m_b)} \pm 0.024 \mathrm{(1/m_b^3)}) 
\end{displaymath}
can be used to extract $|V_{cb}|$~\cite{lephfs}. 
By taking the inclusive and charmless s.l.\ branching fractions presented 
above 
and the average $b$~lifetime $\tau_b$ = 1.564 $\pm$ 0.014~ps, the result is 
$|V_{cb}|$ = (40.7 $\pm$ 0.5~(exp) $\pm$ 2.0~(th)) $\times 10^{-3}$. This 
inclusive method is limited 
by the lepton spectrum model systematics from the s.l.\ branching ratio and by 
the theory systematics due to the values of $m_b$ and $\mu_{\pi}^2$ 
(or $-\lambda_1$) and to the
use of only the first terms of the operator product expansion as highlighted 
in Eq.~(2).

In the exclusive method, the quantity  $|V_{cb}| \times F(1)$ is determined 
from the differential decay rate extrapolated to zero recoil. 
The extraction of $|V_{cb}|$ relies on heavy quark symmetry for the form 
factor normalization $F(1) = 1$ in the heavy quark limit and requires the 
computation of the finite $b$-quark mass effect and of non-perturbative
QCD corrections. The value $F(1) = 0.880 - 0.024 \frac{\mu^2_{\pi} -0.5}{0.1}
\pm 0.035~({\mathrm{excit}}) \pm 0.010~({\mathrm{pert}}) 
\pm 0.025~(1/m^3_b)$ has been adopted by the {\sc Lep} working 
group~\cite{lephfs} giving $|V_{cb}|$ = (39.8 $\pm$ 1.8~(exp) $\pm$ 
2.2~(th)) $\times 10^{-3}$. This result is in agreement with that 
obtained with the inclusive method. There has been a recent lattice 
determination of $F(1)$; unquenched computations will provide smaller and 
better understood systematics. However the larger value obtained by the
recent {\sc Cleo} analysis suggests a cautious attitude and the need for more 
experimental data.
 
\section{OPEN QUESTIONS AND\\ MODEL SYSTEMATICS}

As more data on s.l.\ $b$ decays have become available and the analysis 
techniques improved, the determinations of basic parameters of 
$b$~decays, such as the $|V_{ub}|$ and $|V_{cb}|$ elements, are limited by 
theoretical uncertainties in their derivation from the experimental 
observables and by models in the description of the signal and 
backgrounds~\cite{workshop}. 
While this is promoting new phenomenological approaches, it also requires new 
measurements to better determine the input parameters and further constrain 
the models. This interplay between progresses in theory and new experimental 
insights addresses three main classes of questions: 
i) the definition of a coherent method to include bound state effects in the 
description of inclusive decays, ii) the accurate determination of the 
fundamental parameters, $m_b$, $m_b - m_c$ and $p^2_b$ (or $-\lambda_1$), and 
iii) the estimate 
of the effects of the missing terms in the OPE and of violations of
quark-hadron duality. The present experimental data and the theory predictions
have already significantly progressed towards answers to these questions.

At present, inclusive spectra are predicted by a variety of specialized models
ranging from fully inclusive models, such as the ACCMM model~\cite{accmm}, to 
those saturating the inclusive decay width by the contribution of several 
exclusive final states, like the ISGW and ISGW-2 models~\cite{isgw2}. 

\begin{figure}[h!]

\vspace{-0.5cm}

\begin{center}
\epsfig{file=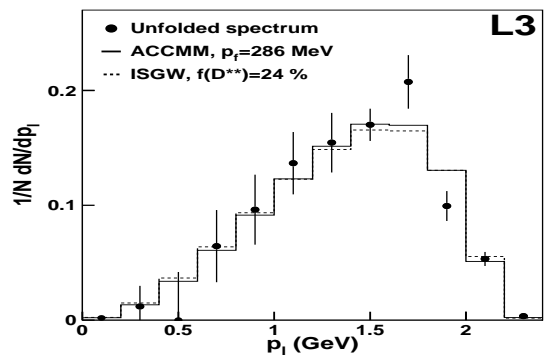,width=8.0cm,height=5.5cm,clip=}
\end{center}

\vspace{-1.25cm}

\caption{\sl The lepton energy distribution, unfolded for detector effects, 
measured by L3~\cite{l3_pf} with overlayed the spectra predicted by the ACCMM 
model and the ISGW model with 24\% of $D^{**}$ contribution}
\label{fig:l3shape}

\vspace{-0.5cm}

\end{figure}
While these models can describe the data quite precisely, after tuning of 
their parameters, their application {\it tout court} to different processes 
is often not justifiable. In inclusive models, the spectra are obtained from a
free quark decaying into partons 
and the non-perturbative are corrections included by convoluting
the parton spectra with a function encoding the kinematics of the $b$ and 
spectator $\bar q$ quarks inside the heavy hadron~\cite{defazio}. 
The ACCMM model describes the $B$ 
hadron as consisting of the $b$ and spectator $\bar q$ quarks, moving 
back-to-back in the $B$ rest frame, with a momentum $p$ distributed according 
to a Gaussian distribution. The Fermi motion width $p_F$ represents a free 
parameter in the model. The value of $p_F$ is proportional to the average 
kinetic energy of the $b$ quark in the hadron $<p_b^2> = \frac{3}{2} p_F^2$ 
and can be derived from the shapes of inclusive observables (see 
Figure~\ref{fig:l3shape}). 

\begin{figure}[ht!]

\vspace{-0.5cm}

\begin{center}
\epsfig{file=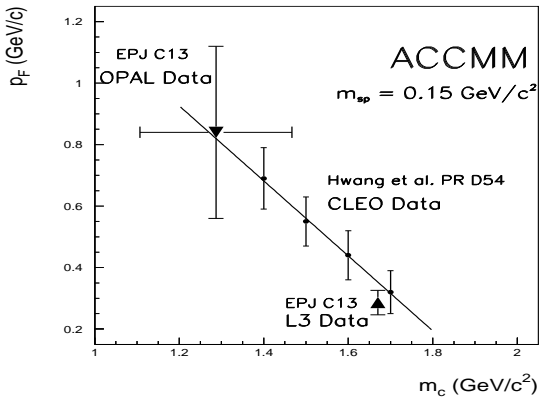,width=8.0cm,height=6.00cm,clip=}
\end{center}

\vspace{-1.25cm}

\caption{\sl Fermi motion $p_F$ vs. charm mass $m_c$ obtained from fits to 
lepton spectra for the ACCMM model. The fit to {\sc Cleo} data was repeated 
for different fixed values of $m_c$~\cite{hwang}, L3 used a single $m_c$ 
value~\cite{l3_pf} while {\sc Opal} performed a two-parameter 
fit~\cite{opal_pf}.}
\label{fig:pffit}

\vspace{-0.5cm}

\end{figure}
Figure~\ref{fig:pffit} summarizes the results from fits to the energy spectrum
of the lepton in s.l.\ $b \rightarrow X \ell \bar \nu$ decays from {\sc Lep} 
and {\sc Cleo} data, showing a good consistency. 
The main issue here is to establish the validity of the ACCMM model, 
with these fitted values, to other decays, such as $b \rightarrow X_u \ell 
\bar \nu$ and $b \rightarrow s \gamma$.

This difficulty may be overcome by a description of the Fermi motion in 
the framework of QCD.  This introduces a shape function, $f(k_+)$,  
corresponding to the distribution of the light-cone residual momentum of the 
heavy quark inside the hadron~\cite{neubert2,dsu,aglietti}. 
At leading order and in the large $m_b$ limit, the light-cone residual 
momentum $k_+$ can be expressed as the difference between the $b$ quark 
mass and its effective mass $m^*_b$ inside the hadron: $m^*_b = m_b + k_+$. 
However, the functional form of the function $f(k_+)$ is not {\it a priori} 
known, except for its first three moments.
The effect of different {\it ansatz} for the description of the 
Fermi motion of the $b$~quark inside the heavy hadron has been studied in the 
case of $b \rightarrow X_u \ell \bar \nu$ both at the parton 
level~\cite{falk,btool} and for the physical observables after full detector 
simulation~\cite{aleph_vub,delphi_vub}. It was found that the Fermi motion 
description contributes with an uncertainty on the branching fraction 
determination of $\pm$~5 - 15\%, increasing as the hadronic mass $M_X$ cut to 
distinguish $b \rightarrow u$ from $b \rightarrow c$ transitions is lowered 
and depending also on the other selection criteria. 

Since $m_b$ enters at the fifth power in the expression of the decay width,
$\Gamma \propto m_b^5 |V_{CKM}|^2$, it is important to determine its value 
with an accuracy of better than $\pm$ 100~MeV/$c^2$ to guarantee a few 
percent error contribution in the extraction of $|V_{ub}|$ and $|V_{cb}|$ 
from inclusive decays. The dependence of the shape of inclusive spectra
on the $b$~quark mass represents a further source of systematics from $m_b$. 
Recent estimates 
from $\Upsilon$ spectroscopy, QCD sum rules and NNLO unquenched lattice 
computations have shown a remarkable agreement~\cite{m_b_rev}. 
On the basis of the 
results~\cite{m_b}, the {\sc Lep} working groups have adopted 
$m_b (1 {\mathrm{GeV}})$ = (4.58 $\pm$ 0.06)~GeV/$c^2$, where 
$m_b (\mu)$ is defined such as $\frac{d m_b (\mu)}{d \mu} = - \frac{16}{9}
\frac{\alpha_s (\mu)}{\pi} + ...$ and the quoted uncertainty defines a 
68\% confidence region. The size of this error is still a topic of some 
controversy and its definition will have a significant impact on the accuracy 
of the determination of $|V_{ub}|$ and $|V_{cb}|$. 

These uncertainties could be reduced by extracting information on the shape 
function and the quark masses from the s.l.\ decay spectra themselves. Due to 
its high statistics, the inclusive $b \rightarrow X_c \ell \bar \nu$ decay is 
the obvious candidate. Since the shape function is expected to be universal 
in $B$ decays, these data could in turn be compared with that extracted from 
other decays and used to constrain models and parameters for 
inclusive, rare decays such as $b \rightarrow X_u \ell \bar \nu$ and 
$b \rightarrow s \gamma$. This analysis appears to be most advantageous for 
experiments at the $\Upsilon(4S)$ since the small $B$ boost minimizes the 
finite resolution effects in converting from the lab frame to the $B$ rest 
frame. On the other hand the lower energy cut-off for lepton identification, 
at $E_{\ell} >$ 0.6~GeV/c, introduces a model dependence when extrapolating 
to zero lepton energy. 
The first moments of the lepton energy and hadronic mass spectra can be 
used to determine $\bar{\Lambda} = m_B - m_b + ...$ and 
$\lambda_1$~\cite{moments}. An analysis of the {\sc Cleo} lepton energy
spectrum limited to the region $E_{\ell} >$ 1.5~GeV gave $\bar{\Lambda}$ = 
(0.39 $\pm$ 0.11)~GeV and $-\lambda_1$ = (0.19 $\pm$ 
0.10)~GeV$^2$~\cite{ligeti_moments}.
{\sc Cleo} reported the preliminary results from the first combined study of 
the first two moments of the hadronic mass and the lepton energy 
spectra~\cite{cleo_moments}. The hadronic mass moments were found to be in 
good agreement with the previous result, while the lepton 
energy spectrum gave unlikely and incompatible values. However, the model 
dependence in the extrapolation to the full spectrum and the unknown 
contributions of higher order terms $1/m^3$ prevent the derivation of any 
conclusions from these disagreements and highlight the importance of further
data. A first lepton energy spectrum for $B \rightarrow X_u \ell \bar 
\nu$ has been extracted by the {\sc Delphi} analysis after unfolding of the 
detector effects. The available statistics is still too small for a 
significant test that may become possible after combining with the data from 
the $B$~factories.  

The predictions for the inclusive analyses presented before and the extraction
of the CKM matrix elements both rely on the basic assumption of duality, 
i.e. that the rates computed at the parton level correspond to those for
the physical final states, after integrating enough hadronic channels. 
The validity of this assumption has been studied in a QCD 
model at the large $N_c$ and small velocity limit~\cite{aleksan} and in the 
(1+1) dimension 't~Hooft model~\cite{thooft}. It was found that the 
sum over exclusive channels corresponds to the inclusive OPE prediction to a 
very good accuracy soon after having integrated the first resonant states.
It is interesting to comment here on the consistency of the determinations of
$|V_{ub}|$ and $|V_{cb}|$ with inclusive and exclusive methods discussed
above. The agreement between the measured values is a test of the 
accuracy of the theoretical methods and also of the validity of the 
quark-hadron duality assumption, up to the
combined measurement uncertainties, i.e. $\simeq$ 8\% for $b \rightarrow c$ 
and 25\% for $b \rightarrow u$. With improved analyses and computational 
techniques and using the larger data sets, already becoming available to 
{\sc Cleo III}, {\sc BaBar} and {\sc Belle}, these tests may reach a 
sensitivity of $\simeq 10\%$ within a few years. Further tests of 
quark-hadron duality that have been proposed include the study of 
$D - \bar D$ mixing and the determination of $V_{ub}$ and $V_{cb}$ for 
$B^0_s$ and $B^{0}_d + B^{-}_u$ separately by an analysis of 
$B^0_s \rightarrow X_u \ell \bar \nu$ and $B^0_s \rightarrow D^*_s
\ell \bar \nu$ decays. 
{\sc Focus}~\cite{focus} and {\sc Cleo}~\cite{cleo_d} have 
already presented new results on neutral $D^0$ decays with enhanced 
sensitivity. The $B^0_s$ meson may further contribute to the investigation 
of quark-hadron duality at a future $e^+e^-$ linear collider able to deliver 
$10^9$ $Z^0$ events, i.e. about 7~M s.l.~$B^0_s$ decays, continuing the 
complementary roles of low- and high-energy colliders in the understanding of 
$B$ decays.

\vspace{0.30cm}

\noindent
{\bf ACKNOWLEDGEMENTS}

\vspace{0.30cm}

I wish to express my gratitude to U.~Aglietti, I.~Bigi, S.~Blyth, G.~Buchalla, 
G.~Martinelli, P.~Roudeau, N.~Uraltsev and W. Venus for their suggestions. 
The {\sc Lep} results presented in this review have been provided by the LEP 
Working Groups on Electroweak Physics, $B$~lifetimes, $V_{ub}$ and $V_{cb}$.
I am also grateful to the organisers of the QCD00 Euroconference for their
invitation and the inspiring atmosphere in Montpellier.

\end{document}